RESEARCH ARTICLE

# Information integration in large brain networks

Daniel Toker 🄳*, Friedrich T. Sommer

Helen Wills Neuroscience Institute, University of California Berkeley, Berkeley, California, United States of America

* danieltoker@berkeley.edu

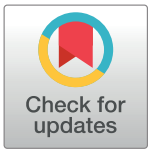

**Data Availability Statement:** The ECoG data used in this paper can be found on NeuroTycho.org, which is a publicly available repository. Some sample simulation data is also included as part of the toolbox linked to in the paper.

**Funding:** This work was supported by the National Science Foundation Graduate Research Fellowship Program under Grant no. DGE 1106400. The funders had no role in study design, data collection and analysis, decision to publish, or preparation of the manuscript.

**Competing interests:** The authors have declared that no competing interests exist.

## Abstract

An outstanding problem in neuroscience is to understand how information is integrated across the many modules of the brain. While classic information-theoretic measures have transformed our understanding of feedforward information processing in the brain's sensory periphery, comparable measures for information flow in the massively recurrent networks of the rest of the brain have been lacking. To address this, recent work in information theory has produced a sound measure of network-wide "integrated information", which can be estimated from time-series data. But, a computational hurdle has stymied attempts to measure large-scale information integration in real brains. Specifically, the measurement of integrated information involves a combinatorial search for the informational "weakest link" of a network, a process whose computation time explodes super-exponentially with network size. Here, we show that spectral clustering, applied on the correlation matrix of time-series data, provides an approximate but robust solution to the search for the the informational weakest link of large networks. This reduces the computation time for integrated information in large systems from longer than the lifespan of the universe to just minutes. We evaluate this solution in brain-like systems of coupled oscillators as well as in high-density electrocorticography data from two macaque monkeys, and show that the informational "weakest link" of the monkey cortex splits posterior sensory areas from anterior association areas. Finally, we use our solution to provide evidence in support of the long-standing hypothesis that information integration is maximized by networks with a high global efficiency, and that modular network structures promote the segregation of information.

## Author summary

Information theory has been key to our understanding of the feedforward pathways of the brain's sensory periphery. But, traditional information-theoretic measures only quantify communication between *pairs* of transmitters and receivers, and have been of limited utility in decoding signals in the recurrent *networks* that dominate the rest of the brain. To address this shortcoming, a theoretically sound measure of information integration has recently been derived, which can quantify communication across an entire brain network. This measure could be pivotal in understanding recurrent brain networks. But, a computational hurdle has made it impossible to quantify this measure in real brains. We present an approximate but robust solution to this hurdle, and use our solution to test long-held assumptions about how brain networks might integrate information.

## Introduction

Information theory, which largely measures communication between transmitter-receiver *pairs* (for e.g. a telephone sender and receiver) [1], has been key to understanding information transmission in the feedforward paths of the brain's sensory periphery [2–8]. But, traditional information-theoretic measures are of limited utility as soon as signals enter the recurrent networks that form the rest of the brain. That is because these measures are designed to quantify *feedforward* information flow. Until very recently, no theoretically sound measures were available to quantify and analyze information that is integrated by entire recurrent *networks*.

Recent work in information theory has risen to meet the challenge of quantifying the integration of information across the recurrent networks that bridge spatially distributed brain areas. Over the last decade, several measures of network-wide information integration have been proposed [9–16], which all generally define information integration as how much more information flows in a *whole* network than in the sum of its parts. The intuition can be phrased like this: if you cut a network into disconnected parts, forcing those parts to evolve over time independently of one another, how much less information is carried over time in the network? If we can estimate this difference accurately, we'd have a value—in bits—of how much information is integrated in a network.

Most of these measures of information integration have faced serious theoretical issues, such as exceeding the total information in a network, falling below 0 bits, or being impossible to estimate from time-series data [11]. To remedy this problem, mathematicians have recently derived a new, theoretically sound measure of information integration called "geometric integrated information", which is immune to the criticisms leveled against most previous measures [17, 18] (that said, we note that a mathematically similar measure called "stochastic interaction" was derived almost two decades ago [9], and that its time-reverse equivalent was recently lauded as a theoretically sound option for measuring information integration [11], but that this measure has been shown to exceed a system's total mutual information in time [14]—a criticism to which geometric integrated information is immune. We also note that there might be other sensible upper-bounds for a measure of integrated information, such as channel capacity or "effective information", as in [19]). This means that, in principle, neuroscientists could use geometric integrated information to push past the feedforward circuits of the brain's sensory periphery, and begin to make sense of the information being integrated across the recurrently connected modules of the rest of the brain.

But there's a hitch. Calculating any of the proposed measures of information integration, including geometric integrated information, is computationally intractable for networks with more than about 20 nodes (e.g. 20 neurons or voxels). That is because all such measures of information integration require identifying what is called the "minimum information bipartition" (MIB) of a network, which is the bipartition that splits the network into two maximally independent sub-communities [9–18]. This makes measuring integrated information in large networks impossible, because finding the MIB requires a brute-force search through all possible bipartitions of a network—a combinatorial search whose computation time explodes super-exponentially with network size.

The reason we need to find the MIB is that a network's capacity for information integration is characterized by where information integration is lowest, which is very much like defining the strength of a chain by the strength of its weakest link: if one link is weak, then the whole chain is weak. For example, if a network has unconnected sub-networks, then the integrated information of that network is 0 bits. In general, to accurately determine a network's value of integrated information, one has to find the MIB of that network. Note that, in principle, the partition that yields *the* global minimum of integrated information might split a network into more than two sub-communities. But, because the number of possible n-partitions explodes with the Bell number (e.g. a network of 8 nodes can be partitioned 4,140 ways, a network of 10 nodes can be partitioned 115,975 ways, and a network of 12 nodes can be partitioned 4,213,597 ways), we follow most of the Integrated Information Theory literature [9–18] and restrict partitions to *bi*partitions, which still capture a network's overall capacity for information integration, and are at least computationally tractable for small networks. But, even with the restriction to bipartitions, the application of Integrated Information Theory is computationally challenging. As mentioned above, a brute-force search to find the bipartition that minimizes integrated information becomes computationally intractable quickly (e.g. a 20-node network can be bipartitioned 524,287 ways and a 30-node network can be bipartitioned 536,870,911 ways). Given the computational intractability of finding the MIB of large networks, our question is this: for a given set of time-series data recorded from nodes in a connected network, is there a way to approximate the minimum information bipartition without a brute-force search?

There have been several proposed solutions to this problem. In our own earlier work [20], we proposed using graph clustering to quickly find the MIB—a proposal also voiced by others [11]—though neither we nor others have yet successfully demonstrated that graph clustering does in fact find good partitions across which to calculate integrated information. Other proposed solutions have used optimization algorithms to find the MIB [21], but these are either prohibitively slow or split brain networks into one-vs-all partitions, which do not reflect how complex biological systems are likely organized [19, 22]. Here, we build upon and empirically validate our earlier proposal that the MIB can be identified through graph clustering.

We show that a network partitioning method called "spectral clustering" [23, 24], when applied to correlation matrices of neural time-series data (Fig 1), reliably identifies or approximates the MIB of even large systems. We demonstrate this in several steps. First, we show that spectral clustering can find the exact MIB in small, brain-like networks (14-16 nodes) of coupled oscillators. Then, we move onto large networks of coupled oscillators (50-300 nodes), where we forced the MIB onto the networks by structurally severing them in half, and show that spectral clustering can find good approximations of the MIB in these large oscillator networks as well. Third, we show that spectral clustering can find the exact MIB in small samples of monkey ECoG data. Fourth, we apply spectral clustering to data from all available recording sites in two monkey brains—which are so large that it would likely take centuries to determine their ground-truth MIB—and show that spectral clustering quickly finds a partition across which integrated information is smaller than or nearly equivalent to the value of integrated information across partitions identified by an optimization-based solution to this search problem (which can take weeks to run).

We note that we also tried using two other community detection algorithms, namely the Weighted Stochastic Block Model algorithm [25] and the Louvain Algorithm for modularity maximization [26], but that our early experimentation with these algorithms did not yield results nearly as strong as did spectral clustering in identifying the MIB. That said, we leave open the possibility that other community detection algorithms might approximate networks' MIBs as well as spectral clustering does.

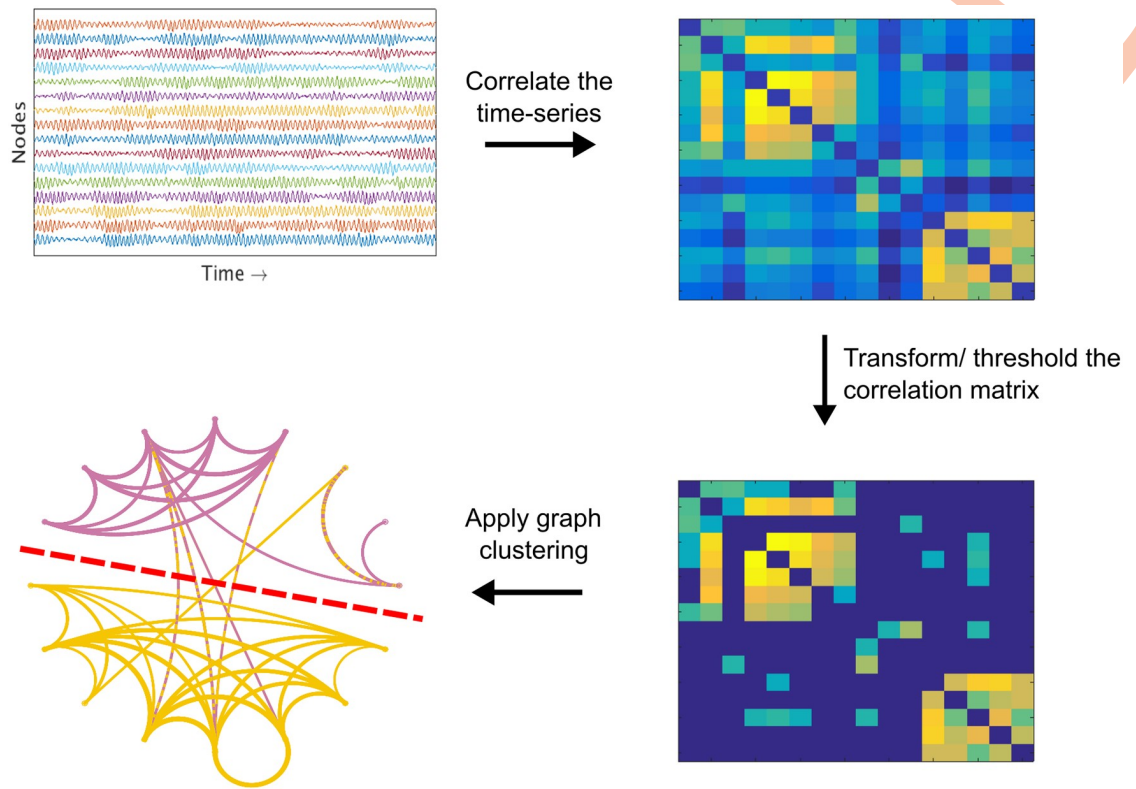

**Fig 1. This is a summary of our method for approximating the minimum information bipartition (MIB) of large systems, which is necessary for calculating integrated information, without a brute-force search.** We assume that the MIB of a brain network is not random, but instead is delineated by the network's functional architecture. To identify the functional architecture of brain networks from time-series data, we draw on work from functional brain connectomics, in which "functional brain networks" are often constructed by taking correlation matrices of neural time-series data, thresholding those correlation matrices to produce weighted adjacency matrices, and applying community detection algorithms like spectral clustering to those adjacency matrices. This procedure partitions the brain into functionally distinct sub-networks [35]. Our hypothesis is that the MIB of a brain network should be delineated by the functional boundaries identified through graph clustering. Out of the range of approaches to clustering brain networks, we chose spectral clustering because it is particularly well-suited for *normalized* partitioning problems, in which (just as with the search for the MIB), the goal is to find sub-networks of roughly equal size (i.e., to avoid partitioning a network into one node isolated from the rest of the network). see Methods for details on how spectral clustering was used to approximate the minimum information bipartition of brain networks.

We use our spectral clustering-based method to report two novel empirical findings: 1) The MIB of ECoG recordings in the macaque cortex splits posterior sensory areas from anterior association areas, and 2) Supporting predictions from neural connectomics research, we show that networks with a high global efficiency (i.e. a short average path length) produce high integrated information and that strongly modular networks produce low integrated information.

Because we believe that this measure will be empirically valuable for understanding how different brain states or task conditions rely on different modes of information integration between neurons or brain regions, we have made our code publicly available as a toolbox at https://figshare.com/articles/Information_Integration_in_Large_Brain_Networks/7176557.

## Results

### Geometric integrated information

As mentioned in the Introduction, a number of measures of integrated information based on time-series data have been proposed. Only very recently [17, 18], a measure was derived that is at the same time computable from time-series data and properly bounded between zero bits and the total mutual information in time and space in a system. This measure, called "geometric integrated information", or $\Phi^G$, is defined as the minimized Kullback-Leibler divergence between the "full model" $p$ of a system $X$, which fully characterizes all the spatiotemporal influences within the system, and a "disconnected model" $q$. In the disconnected model, the network of interest is partitioned into statistically disconnected sub-communities, which evolve over time independently of one another:

$$q(X_t^i | X_{t-\tau}) = q(X_t^i | X_{t-\tau}^i) \; \forall i \tag{1}$$

where the index $i$ labels the statistically disconnected sub-communities (so, for a bipartition, $i$ iterates from 1 to 2), and $X_t$ and $X_{t-\tau}$ describe present and past states of the system, respectively ($t$ and $t-\tau$ are discrete time indices). $X_t^i$ and $X_{t-\tau}^i$ refer to non-empty subsets (corresponding to sub-communities) of the variables constituting $X_t$ and $X_{t-\tau}$; $X_t$ and $X_{t-\tau}$ are $n$-dimensional real-valued random vectors, i.e. $X_t := (X_{t_1}, X_{t_2}, ..., X_{t_n})$, where $X_{t_j}$ for $j = (1, ..., n)$ are real-valued random variables. In other words, for a given multivariate time-series, with $n$ variables (e.g. neurons, electrodes, or voxels) and $m$ time-points, $X_{t-\tau}$ is a matrix of observations of all $n$ variables from time 1 to time $m-\tau$, and $X_t$ is a matrix of observations of al $n$ variables from time $\tau$ to time $m$. Geometric integrated information is then defined as:

$$\Phi^G = \min_q D_{KL}[p(X_t, X_{t-\tau}) || q(X_t, X_{t-\tau})] \tag{2}$$

where $D_{KL}[p, q]$ stands for the Kullback-Leibler divergence between two distributions $p$ and $q$.

Geometric integrated information has a simple and quick-to-compute formulation for multivariate Gaussian signals [17], and all data analyzed in this paper are approximately multivariate normal (S2 Fig). (We note that for Gaussian variables no recourse to information geometry is necessary to minimize the KL divergence in Eq 2, and so arguably there is no direct sense in which this measure is "geometric" for Gaussian variables. That said, because the framework of information geometry is necessary for calculation of this measure in the non-Gaussian case, we follow [17] and still call this measure "geometric integrated information" in the Gaussian case). Like many information-theoretic measures, geometric integrated information can be computed in Gaussian data using the framework of linear regression. As is commonly done in time-series analysis across a range of fields, we can model the evolution in time of a Gaussian system using a simple linear regression model:

$$X_t = AX_{t-\tau} + E \tag{3}$$

where $X_t$ corresponds to the present of the system and $X_{t-\tau}$ corresponds to the past of the system, $A$ corresponds to the regression matrix estimated from the data, and $E$ corresponds to the error or residuals in the linear regression. Both $A$ and $E$ can be computed from the covariance matrices of the data. The regression matrix $A$ is given by the normal equation:

$$A = \Sigma_{X_t X_{t-\tau}} (\Sigma_{X_{t-\tau} X_{t-\tau}})^{-1} \tag{4}$$

where $\Sigma_{X_t X_{t-\tau}}$ is the variance between the present and the past of the system $X$. The covariance of the error matrix $E$ can also be computed from the covariance of the data, and is precisely equivalent to the *conditional variance* of the present, given the past of the system:

$$\Sigma_{EE} = \Sigma_{X_t | X_{t-\tau}} \tag{5}$$

where

$$\Sigma_{X_t | X_{t-\tau}} = \Sigma_{X_{t-\tau} X_{t-\tau}} - A(\Sigma_{X_t X_{t-\tau}})^T \tag{6}$$

The covariance of $E$ is all we'll need for the *complete* model of the system's evolution. Oizumi et al [17] prove that the disconnected model of the system can also be expressed in terms of linear regression:

$$X_t = A'X_{t-\tau} + E' \tag{7}$$

where $A'$ is a regression matrix like $A$, but all elements describing interactions across the MIB have been set to zero (i.e. $A'$ is a diagonal block matrix). If we have correctly identified the MIB of the network, and therefore set all the right elements of $A'$ to zero, then the covariance of $E'$, which is the only thing we now need to calculate integrated information, is:

$$\Sigma_{E'E'} = \Sigma_{EE} + (A - A')\Sigma_{X_{t-\tau} X_{t-\tau}}(A - A')^T \tag{8}$$

There is no (known) closed-form solution for $A'$ and $\Sigma_{E'E'}$, but these matrices can be estimated using iterative methods. In this paper, we estimate these matrices using the augmented Lagrangian method provided by [27]. Finally, we insert Eqs 6 and 8 into the standard formula for the Kullback-Leibler divergence between two Gaussians with identical means. After a simple algebraic transformation the estimate of integrated information, in bits, can be written:

$$\Phi^G = \frac{1}{2}\log\frac{|\Sigma_{E'E'}|}{|\Sigma_{EE}|} \tag{9}$$

where $|\Sigma_{E'E'}|$ refers to the determinant of the error matrix in our disconnected model, and $|\Sigma_{EE}|$ refers to the determinant of the error matrix in the connected model.

If the sub-communities of a network evolve in time mostly independently of one another, then these determinants will be close and $\Phi^G$ will be small. If, on the other hand, there are strong inter-dependencies between the sub-communities of a network, then these two determinants will diverge and $\Phi^G$ will be large.

To find the minimum information bipartition, we need to perform a brute-force search through all possible bipartitions of a network, and find the bipartition that minimizes integrated information. Unfortunately, this will usually lead to strongly asymmetrical partitions, in which one or two nodes are split from the rest of the system—and such partitions are usually of little functional relevance [11, 14–16]. While how to best handle such asymmetric partitions remains an open problem in the Integrated Information Theory literature [11, 14], there have been a number of proposed solutions for finding more balanced and functionally meaningful partitions. Here, we use the solution originally suggested in [19] and also used in [15, 16], which is to find the bipartition that minimizes integrated information, normalized by the factor $K$:

$$K = \min_k [H(M^k)] \tag{10}$$

where $H(M^k)$ refers to the entropy of a sub-community $M^k$. For a multivariate Gaussian system $M$, the entropy $H(M) = \frac{1}{2}\ln(|2\pi e \Sigma(M)|)$, where the bars denote the matrix determinant and $\Sigma(M)$ is the covariance matrix of the variable $M$. Normalized integrated information thus equals $\frac{\Phi^G}{K}$. Minimizing the *normalized* version of integrated information biases the search toward partitions that are more balanced in the number of nodes, and away from partitions in which a single node is isolated from the rest of the network. Thus, strictly speaking, the MIB of a network is the bipartition, out of all possible bipartitions, that minimizes $\frac{\Phi^G}{K}$, and the integrated information of that network is $\Phi^G$, *not* normalized by $K$, across that partition. (Note that normalization was not discussed in the paper in which geometric integrated information was originally derived [17], but that it has already been shown that without normalization, the bipartition that minimizes geometric integrated information is often the one-vs-all partition [21]).

Recall that earlier, we mentioned that a previously proposed optimization-based solution for quickly finding the MIB often splits networks into one-vs-all partitions, which are difficult to interpret in terms of biological function. This solution, proposed by [21], makes use of the Queyranne algorithm for minimizing sub-modular functions, and was shown to accurately identify bipartitions that minimize *non*-normalized integrated information. Problematically, these bipartitions are often one-vs-all splits—which is precisely what normalization was designed to avoid. Thus, finding the MIB using the Queyranne algorithm can be considered a valid option if a researcher wants to find a partition that minimizes non-normalized integrated information, as opposed to normalized integrated information. Our goal, however, is to find a quick and accurate method for identifying bipartitions that minimize *normalized* integrated informaton, because we share others' conviction [19] that this yields more biologically meaningful results.

Finally, note that $\Phi^G$ is calculated over a time-lag $\tau$ (Eqs 1–8). If, for example, $\tau$ is set to 50 ms, then $\Phi^G$ will tell you, in bits, how much information is carried over 50 ms using the network connections that cross the MIB of your system. While the choice of a partition across which to calculate integrated information (i.e., the MIB) is well-defined, the choice of a time-lag $\tau$ is not. For the purposes of this study, we chose a time-lag that, on average, maximized integrated information for the system at hand (S3 Fig). This choice was based on previous papers [13, 28], which, based on phenomenological arguments, maintain that the time-scale of neural information integration that is most relevant to cognitive and perceptual processes is the scale that maximizes integrated information—a claim about which we are agnostic, but which our method could help elucidate in future research. That said, we note that in general, it is common to estimate time-delayed information measures such as transfer entropy for various time-lags, and then to choose the time-lag that maximizes the information measure of interest. This procedure has been shown to accurately capture the time scales of delayed system interactions [29].

## Identifying the MIB with graph clustering

As a critical innovation, which enables the estimation of $\Phi^G$ for large networks, we propose to reduce the search space for the MIB using graph clustering on the correlation matrix of neural time-series data (Fig 1). We searched the literature for a graph clustering algorithm that is biased toward balanced partitions, like the search for the MIB. We therefore chose to use spectral clustering [23] to partition our networks, because it is known to quickly find bipartitions that approximately but robustly minimize the "normalized cut function" in graph theory, which is the sum of weights that cross a partition normalized by the sum of weights between the entire network and the communities on either side of that partition (see Methods for more details). While the normalized cut function is mathematically distinct from the function being minimized in search for the MIB (i.e., $\frac{\Phi^G}{K}$), in both cases normalization is being used to find roughly equal-sized communities, and so we hypothesized that both should yield similar partitions.

To use a network partitioning algorithm, we need a way to estimate network structure from time-series data. To address this challenge, we drew on insights from neural connectomics research. Network neuroscientists often treat the correlation matrix of neural time-series data as a "functional network" describing neural interactions, and apply graph clustering algorithms like spectral clustering to neural correlation matrices to partition the brain into distinct functional sub-networks [22, 30–35]. Following this insight, our method takes the correlation matrix of time-series data, transforms it using a power adjacency function (following [36]) and thresholds the transformed matrix across a range of cutoffs (following [37–41]), applies spectral clustering at each threshold, calculates $\Phi^G$ (normalized) across each resulting candidate network partition, and picks as the estimate of the MIB the partition that yields the lowest value of $\Phi^G$ (normalized). See the Methods for more details on how we used spectral clustering to approximate the MIB.

## Spectral clustering finds the MIB in small brain-like networks of coupled oscillators

As a first step in assessing how well spectral clustering on the correlation matrix of time-series data recorded from a network can find the MIB of that network, we begin with a simulation of coupled oscillators. Among the variety of existing oscillator models, we chose to test our method in brain-like networks of coupled stochastic Rössler oscillators [42] because, when weakly coupled, their activity approximates a multivariate normal distribution [43] (S2A–S2C and S2F–S2K Fig), similar to the ECoG data we analyze later in this paper (S2D and S2E Fig). Besides oscillators' frequency and the amplitude of noise injected into the oscillators, all parameters in the model were taken from previous literature (see Methods).

We simulated 25,000 time-points of oscillatory signals from 50 14-node networks and 50 16-node networks. These networks were generated using a novel algorithm based on Hebbian plasticity, which produces connectivity patterns that recapitulate basic features of brain connectomes, including a modular structure and rich between-module connectivity [22], and a log-normal degree distribution [44] (see Methods).

To assess the performance of spectral clustering in identifying the MIB from time-series data, we need a best guess at the "ground truth" MIB of a system. When the underlying transition probabilities of a system are known, the ground-truth MIB can simply be determined by a brute-force search through all possible bipartitions of a system and identifying the bipartition that minimizes normalized integrated information. Identifying the ground-truth from time-series data, however, requires infinite observations. Thus, when we refer to the "ground-truth" MIB throughout this paper, we simply mean the bipartition, identified through a brute-force search through all possible bipartitions, that minimizes an estimate of normalized integrated information from finite observational data.

We found that in 95/100 of our small simulated networks, there was a difference of 0 bits between $\Phi^G$ (normalized) across the spectral clustering-based bipartition and the lowest value of $\Phi^G$ (normalized) identified through a brute-force search through all possible bipartitions (Fig 2a). In other words, in almost all networks tested, our spectral clustering-based approach gives the exact same result as does a brute-force search for the MIB. We further found that the Rand Index [45] (a common measure of partition similarity) between the ground-truth MIB and the spectral bipartition was 1 (indicating a perfect match) for those same 95 networks (Fig 2b). Finding partitions that are highly similar to the MIB in these networks is important, since the more dissimilar a partition is from the MIB, the larger $\Phi^G$ (normalized) will tend to be across that partition; in other words, the further off you are from the MIB, the less accurate your estimate of integrated information will tend to be (S4A and S4B Fig). To test the statistical stability of these results, we computed running averages of both the Rand Indices and the differences between estimated $\Phi^G$ values (e.g. the running mean Rand Index of the first two 14-node networks, then the first three 14-node networks, then the first four 14-node networks, etc.). We then took the approximate derivatives of the running averages for both network sizes, and used two-sample t-tests to accept the null hypothesis ($\alpha = 0.05$) that the approximate derivatives were indistinguishable from 0 for both tests, for both network sizes. This means that the results reported in Fig 2 are statistically stable at a sample size of 50 networks (i.e. adding more samples would not likely change the means significantly, as the differences in the running average are already approximately zero at just 50 networks). To further check whether this result generalizes across different network dynamics, we used the same networks to generate multivariate autoregressive simulations and performed the exact same analysis, and found that spectral clustering also accurately identifies the MIB for autoregressive data (S7 Fig). We used the same running average and approximate derivative test to confirm that our results for the autoregressive dynamics in S7 Fig are also statistically stable at a sample size of 50 networks.

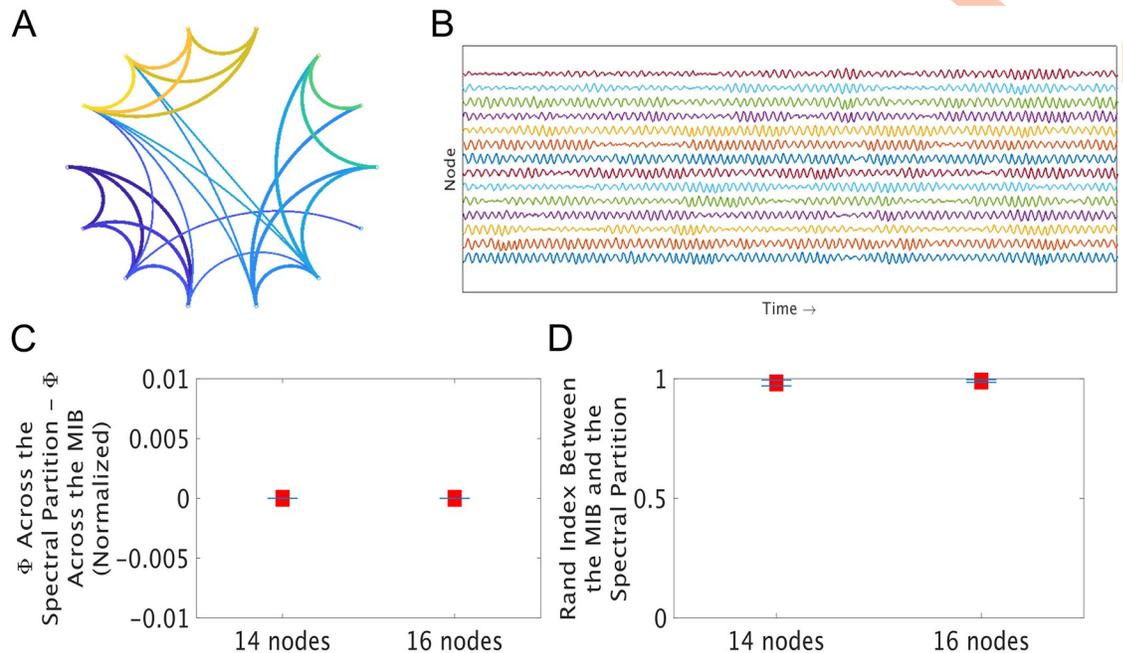

**Fig 2. We first tested our spectral clustering-based approach in small simulations. A** This is an example of a small brain-like network we generated using a novel algorithm based on Hebbian plasticity. This algorithm produces networks that are loosely brain-like, in that they are modular, show rich cross-module connectivity, and display a log-normal degree distribution with long right tails. We used this algorithm to generate 50 14- and 16-node networks. see Methods for more details on network generation. **B** This is a sample of oscillatory data generated from the network in A. We generated these data using a stochastic coupled Rössler oscillator model. In the Rössler oscillator model, each node stochastically oscillates according to its own intrinsic frequency, and dynamically synchronizes with other nodes it is connected to. The resulting data are multivariate normal (S2 Fig), allowing for the fast computation of integrated information. **C** As a first test of our spectral clustering-based approach to identifying the MIB from time-series data, we subtracted $\Phi^G$ (normalized) across the ground-truth MIB, identified through a brute-force search through all possible bipartitions, from $\Phi^G$ (normalized) across the partitions identified through spectral clustering. In this test, a perfect match between values would yield a difference of 0 bits. Red squares indicate the mean across 50 networks, and the blue bars indicate standard error of the mean. **D** As a second test of our spectral clustering-based approach, we computed the Rand Index [45], which is a common measure of partition similarity, between the spectral partitions and the ground-truth MIBs of these networks. A Rand Index of 1 indicates a perfect match between partitions, and a Rand Index of 0 indicates maximum dissimilarity between partitions. Red squares indicate the mean across 50 networks, and the blue bars indicate standard error of the mean. These results show that spectral clustering finds the MIB of small networks of coupled oscillators. We found similar results using the same networks but different network dynamics (S7 Fig).

Finally, we also compared our approach to another proposed method for quickly identifying the MIB from time-series data. This method uses the Queyranne algorithm for fast minimization of sub-modular functions [21]. Though in past work the Queyranne algorithm has been successfully used to minimize non-normalized integrated information, we used the Queyranne algorithm to try to find a bipartition that minimizes *normalized* integrated information. The difference between $\Phi^G$ (normalized) across the Queyranne bipartition and $\Phi^G$ (normalized) across the MIB was 0 bits (indicating a perfect match) in only 1/50 14-node networks (mean difference = 0.0031 bits) and in 2/50 16-node networks (mean difference = 0.0026 bits). The Rand Index between the Queyranne partition and the MIB was 1 for the same networks for which the difference in $\Phi^G$ (normalized) was 0; the mean Rand Index was 0.576 across all 14-node networks and 0.582 across all 16-node networks. The Queyranne algorithm also performed poorly in minimizing normalized integrated information in autoregressive simulations generated from these same small brain-like networks (S7 Fig). Thus, spectral clustering does a better job of estimating the MIB in small brain-like networks than does the Queyranne algorithm. Moreover, even when trying to minimize normalized integrated information, which is biased toward balanced partitions, the Queyranne algorithm often found partitions that isolate one node from the rest of the network. This occurred in 23/50 of the 14-node networks (while none of the MIBs identified through a brute-force search yielded one-vs-all partitions) and in 26/50 of the 16-node networks (while only one of the MIBs identified through a brute-force search was a one-vs-all partition). Such partitions are usually of little functional relevance—hence why normalization is introduced in searching for the MIB [19]. Moreover, the partitions found by the Queyranne algorithm were also generally dissimilar from the partitions found by our spectral clustering approach: the mean Rand index between the spectral partitions and the Queyranne algorithm partitions was 0.57 for the 14-node networks and 0.59 for the 16-node networks.

## Spectral clustering approximates the MIB in large, *Cut* brain-like networks of coupled oscillators

Having passed this basic test in small networks, we next asked whether spectral clustering can accurately identify the MIB in large systems. To test this, we used our algorithm for generating brain-like connectivity (see Methods) to create networks which ranged from 50 to 300 nodes in size. Networks of these sizes cannot be exhaustively searched for their MIB, so we forced the MIB onto these networks by cutting them in half. If a network is cut into two parts, then, with infinite data, the MIB will converge onto where the network has been cut and $\Phi^G$ across this cut will be 0 bits. For these networks, we generated 100,000 time-points of data using the stochastic Rössler oscillator model, since in larger systems more data are necessary for more accurate estimation of multivariate information measures. We were unable to test the accuracy of the Queyranne algorithm for these networks, because the computation time for using the algorithm to minimize normalized integrated information increased exponentially, making its application to networks with more than 50 nodes prohibitively expensive; that said, we note that the algorithm is far faster in minimizing *non*-normalized integrated information, as shown in [21].

Spectral clustering again performed remarkably well. The mean absolute difference between $\Phi^G$ across the spectral partition and $\Phi^G$ across the ground-truth cut was less than 0.001 bits (normalized) for all network sizes (Fig 3A), indicating a close match. Note that, objectively, $\Phi^G$ should be zero in these cut networks, and we would expect estimates of $\Phi^G$ to converge to zero bits with infinite data; as a sanity check, we utilized a well-established method for extrapolating estimates of information measures to what they would be if infinite data were available, and found that this brought estimates significantly closer to zero bits for these cut networks, as expected (S1 Fig). The Rand Index between the spectral partition and the ground-truth cut was greater than .8 for 37/40 of the 50-node networks, 39/40 of the 100-node networks, 29/40 of the 150-node networks, 21/40 of the 200-node networks, 18/40 of the 250-node networks, and 10/40 of the 300-node networks. Given that the estimates of $\Phi^G$ across the spectral partition and the ground-truth cuts were very close even in the 200- to 300-node networks (for which the spectral partitions were similar to the ground-truth cut less often) and also both extrapolated to around the ground-truth of zero bits, these results suggest that there are sometimes multiple minima for normalized integrated information (i.e. in these cut networks, there are sometimes several bipartitions across which there is little to no information integration). To test the statistical stability of these results, we computed running averages of both the Rand Indices and the differences between estimated $\Phi^G$ values (e.g. the running mean Rand Index of the first two 50-node networks, then the first three 50-node networks, then the first four 50-node networks, etc.). We then took the approximate derivatives of the running averages for each network size, and used two-sample t-tests to confirm that the approximate derivatives were statistically indistinguishable from 0 for both tests, for each network size. This means that the results reported in Fig 3 are statistically stable at a sample size of 40 networks (i.e. adding more samples would not likely change the means significantly). Finally, we again checked whether this result generalizes across different network dynamics, by generating autoregresive simulated data from these large, cut networks. We found that spectral clustering performed even better (nearly perfectly) for the autoregressive simulations (S8 Fig), again supporting the robustness and generalizability of our method. We again used a running mean of the results, together with approximate derivatives, to confirm that the results for the autoregressive data in S8 Fig were also statistically stable at a sample size of 40 networks.

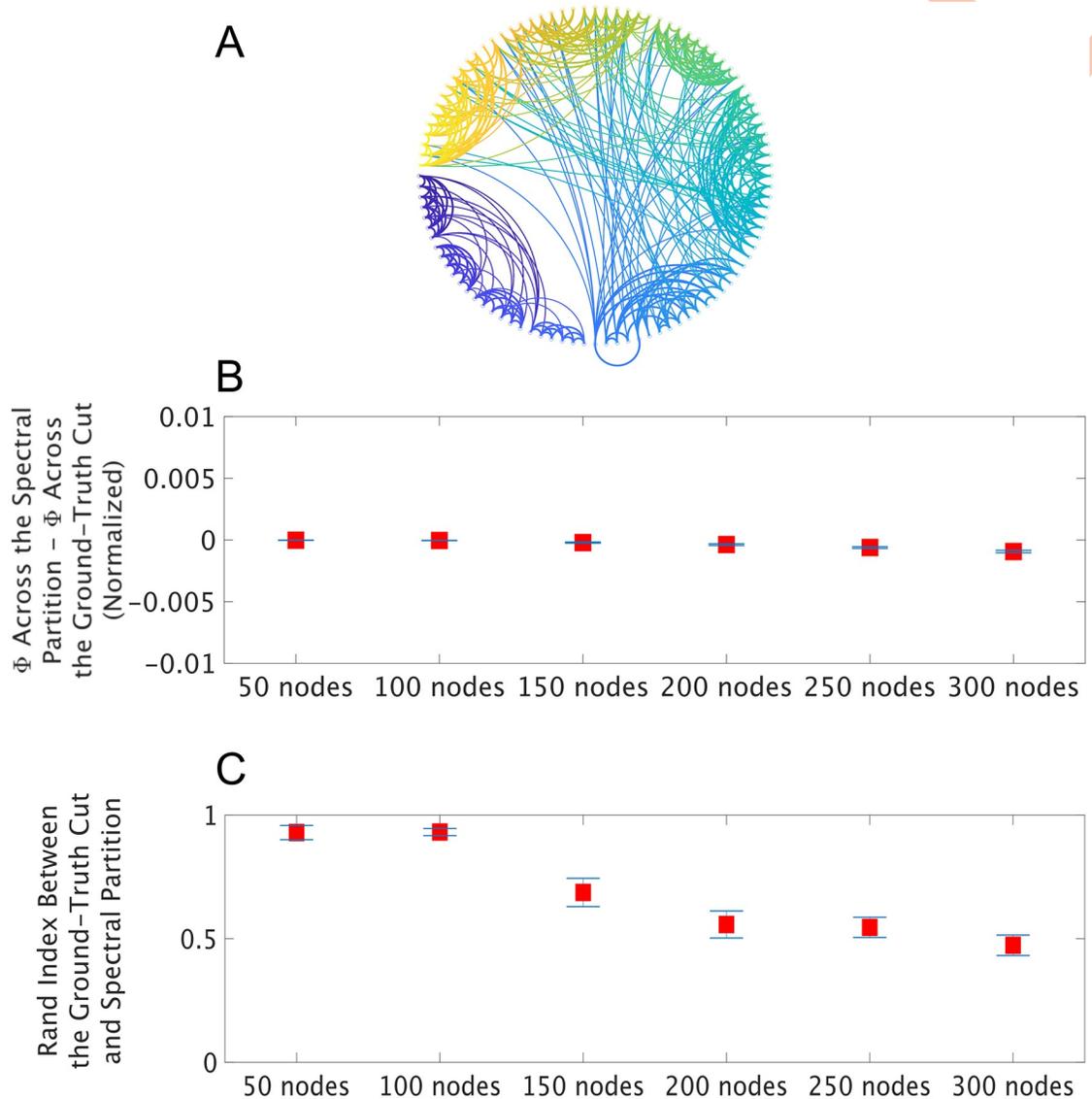

**Fig 3. A** Having shown that spectral clustering can find the MIB in time-series data from small networks, we next asked whether it could find the MIB of large simulated networks. While large networks cannot be exhaustively searched for their MIB, the MIB can be forced onto them by cutting them in half. We generated 40 such cut networks for each network size. Network sizes ranged from 50 nodes to 300 nodes. **B** Here, we show $\Phi^G$ (normalized) across the ground-truth cut subtracted from $\Phi^G$ (normalized) across the partition identified through spectral clustering. Red squares indicate the mean across 40 networks, the absolute value of which never exceeded 0.001 bits (normalized), and the blue bars indicate standard error of the mean. **C** Here, we show the mean and standard error of the Rand Index between the ground-truth cut and the spectral clustering-based partition of the correlation matrix estimated from each network. The Rand Index between the spectral partition and the ground-truth cut was greater than 0.8 (indicating high similarity) for the majority of networks of all network sizes, except for the 200- to 300-node networks. Despite this dip in Rand Index, spectral clustering still found partitions across which $\Phi^G$ (normalized) was extremely close to $\Phi^G$ (normalized) across the ground-truth cut in the 300-node networks (**A**), which suggests that in these networks, there was sometimes several possible partitions that minimized normalized integrated information.

## Spectral clustering approximates the MIB in the macaque cortex

We next applied the same spectral clustering method to one minute of ECoG data from two macaque monkeys, Chibi and George [46]. After pre-processing (see Methods), data for 125 electrodes distributed across the left cortex of each monkey were available. These data were multivariate normal (S2D and S2E Fig). To enable comparison between graph clustering-based partitions and the ground-truth MIB, we divided these data into overlapping sets of fourteen electrodes each, resulting in 112 sets of electrodes for each monkey. The difference between $\Phi^G$ across the MIB and $\Phi^G$ across the partitions identified by spectral clustering was 0 (indicating a perfect match) for 46/112 of the datasets from Chibi's brain (mean difference = 0.0001 bits) and in 67/112 of the datasets from George's brain (mean difference = 0.0002 bits) (Fig 4A). The Rand Index comparing the spectral partition and MIB was 1 for those same datasets (Chibi mean Rand Index = 0.79, George mean Rand Index = 0.87) (Fig 4B). As was the case for our simulated networks, the more dissimilar partitions in the monkeys' brains were from the MIB, the larger $\Phi^G$ (normalized) tended to be across those partitions (S4 Fig). The Queyranne algorithm again performed worse than spectral clustering, yielding perfect matches to the ground-truth in only 18/112 of the datasets from Chibi's brain (mean Rand Index = 0.6) and 22/112 from George's brain (mean Rand Index = 0.64). Moreover, as was the case for our simulated data, the Queyranne algorithm separated one node from the rest of the system in the majority (145/224) of all ECoG datasets (as opposed to the ground-truth MIBs, which separated one node from the rest of the system in only 39/224 datasets). Finally, the partitions found by the two algorithms were generally dissimilar: the mean Rand index between the spectral partitions and the Queyranne algorithm partitions was 0.65 for the electrode clusters in Chibi's brain and 0.67 for George's brain.

As a test of how well spectral clustering could approximate the MIB for all electrodes, we asked whether it could minimize $\Phi^G$ (normalized) in the whole cortex of each monkey. We therefore calculated $\Phi^G$ across the spectral clustering-based bipartition of the entire left cortex for both monkeys. We found that this estimate of the MIB split posterior sensory areas from anterior association areas in both brains (Fig 4C and 4E). To test the statistical robustness of this result, we compared both our estimated $\Phi^G$ (normalized) values and our estimated MIBs

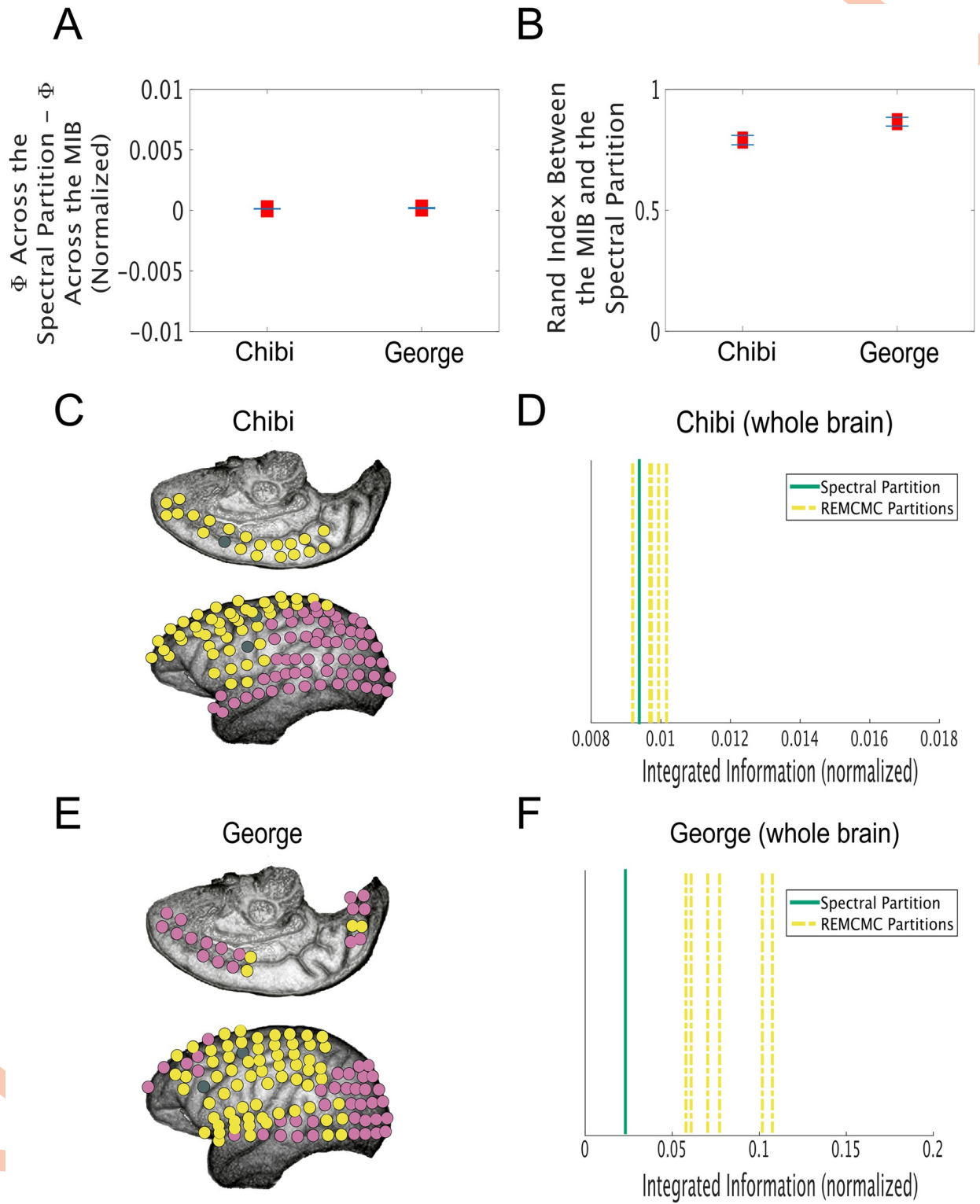

**Fig 4. A** We split the available ECoG electrodes in two macaque monkeys into overlapping sets of 14 electrodes. The ground-truth MIB of 14 electrodes can be identified through a brute-force search, and compared to the spectral partition estimated from the correlation matrix of data from those electrodes. Here, we subtracted $\Phi^G$ (normalized) across the ground-truth MIB from $\Phi^G$ (normalized) across the spectral partition. There was a difference of 0 bits for 67/112 (mean difference = 0.0002 bits) datasets from George's brain, and a difference of 0 bits in 46/112 (mean difference = 0.0001 bits) datasets from Chibi's brain. Red squares indicate the mean difference in $\Phi^G$ (normalized) across all datasets from one

brain, and blue bars indicate standard error of the mean. **B** Spectral clustering found the exact MIB for the same 67/112 datasets in George's brain (mean Rand Index = 0.87) and 46/112 datasets in Chibi's brain (mean Rand Index = 0.79). **C** We used our spectral clustering approach to estimate the MIB of Chibi's entire left cortex, and found that it split posterior sensory areas from anterior association areas. Electrodes are colored according to the community in which they are clustered; the electrodes that were excluded from the analysis because they displayed consistent artifacts are colored grey. **D** $\Phi^G$ (normalized) across the spectral partition of Chibi's left cortex (solid green line) was lower than it was across 4/6 partitions identified by the Replica Exchange Markov Chain Monte Carlo (REMCMC) method (yellow dashed lines) [21]. The other 2/6 partitions yielded values of normalized integrated information that were very slightly lower (0.0002 bits) than the value across the spectral clustering-based partition, and were dissimilar both to each other (Rand Index = 0.5) and to the spectral partition (Rand Indices = 0.5, 0.55), suggesting that there were several local minima of normalized integrated information in Chibi's brain. We ran the algorithm for 10 days. **E** Our estimate of the MIB of George's left cortex using spectral clustering also (largely) split posterior sensory areas from anterior association areas. **F** $\Phi^G$ (normalized) across the spectral partition of George's left cortex was lower than it was across all bipartitions identified by the REMCMC method. Note the difference in scale on the x-axes of **D** and **F**; it is unclear why this scale should differ between the two brains.

for both monkey cortices to results from 100 Amplitude Adjusted Fourier Transform surrogate datasets [47]; we found that our estimated $\Phi^G$ (normalized) values were significantly higher than the surrogate distributions for both monkeys, and that the similarities between the MIBs estimated for the monkey cortices and the MIBs estimated for the surrogate datasets were at chance levels, suggesting that the results for the full monkey brains are not artifactual (S9 Fig). We then compared $\Phi^G$ across the spectral clustering-based partitions to $\Phi^G$ values calculated across partitions identified by a Replica Exchange Markov Chain Monte Carlo (REMCMC) algorithm. The REMCMC method for estimating the MIB is described in detail in [21]; the algorithm used in this paper is the same as that used in [21], except that it searched for a bipartition that minimized normalized (rather than non-normalized) integrated information. We also terminated the algorithm after 10 days, since it failed to reach convergence for either monkey dataset by that point. Since the algorithm tries to minimize normalized integrated information across six parallel sequences, it produces six guesses for the MIB. We also tried using the Queyranne algorithm for the monkey brains, but the algorithm failed to terminate even after two weeks of running, and so we did not include the Queyranne algorithm in this analysis.

For George's brain, normalized integrated information across the spectral clustering-based partition was lower than it was across all six bipartitions identified by the REMCMC method (Fig 4E). In Chibi's brain, the REMCMC algorithm found two partitions across which normalized integrated information was very slightly lower (0.0002 bits) than it was across the spectral clustering-based partition; interestingly, the two REMCMC partitions (which yielded the same value of normalized integrated information) were not only dissimilar to each other (Rand Index = 0.5), but were also both dissimilar to the spectral clustering-based partition (Rand Indices = 0.5, 0.55), suggesting that there were several local minima of normalized integrated information for Chibi's brain. In all, these results show that our spectral clustering-based method reliably minimizes $\Phi^G$ (normalized) of the entire macaque cortex, suggesting that it successfully finds or approximates the MIB in large neural data.

## Network structure and information integration

The ability to quickly measure information integration in large networks allowed us to assess what network architectures best support information integration, and what that might imply about how brains could be organized to integrate information. We here test for the first time, *in silico*, several graph-theoretic measures that have been hypothesized to track neural information integration. Note that in the neural connectomics literature, these graph-theoretic measures are often applied to either *structural* networks, such as the physical connectivity between brain regions that might be revealed through diffusion tractography, or to *functional* networks, such as correlation matrices calculated from functional magnetic resonance imaging recordings [48]. Because analyses of structural networks are more straightforward than analyses of functional networks (primarily because there is considerable debate surrounding what constitutes a functional network), we here focus on the relationships between structural networks and integrated information. We hope to more systematically investigate the relationship between integrated information and *functional* networks in future work.

The most commonly invoked graph-theoretic measure of a network's capacity to integrate information is *global efficiency* [37, 48–51]. Global efficiency is related to the inverse of the average shortest path between nodes in a network. Formally, the global efficiency $E$ of a network $G$ is defined as follows:

$$E(G) = \frac{1}{n(n-1)} \sum_{i \neq j \in G} \frac{1}{d(i,j)} \qquad (11)$$

where $n$ is the number of nodes in the network and $d(i,j)$ is the shortest path between given network nodes $i$ and $j$. In high efficiency networks, any node can be reached by any other node with only a few steps. For about a decade, network neuroscientists have assumed that the global efficiency of a brain network quantifies its ability to concurrently exchange information between its spatially distributed parts; for this reason, it has been assumed that global efficiency sets an upper limit on neural information integration [37, 48, 50, 51].

Conversely, it has been assumed that the modularity of brain networks (and of complex networks more generally) *limits* the integration of information, primarily by segregating network dynamics [22, 33, 51]. The modularity of a network is defined by Newman's $Q$:

$$Q = \frac{1}{2m} \sum_{ij} [A_{ij} - \frac{k_i k_j}{2m}] \delta(c_i, c_j) \qquad (12)$$

where $A_{ij}$ is the adjacency between nodes $i$ and $j$, $k_i$ and $k_j$ are the sums of the adjacencies involving $i$ and $j$, respectively, $c_i$ and $c_j$ are the modules to which nodes $i$ and $j$ have been assigned, respectively, $m = \frac{1}{2} \sum_{ij} A_{ij}$, and $\delta(c_i, c_j)$ equals 1 if $c_i = c_j$ and 0 otherwise. Networks that can be easily subdivided into distinct sub-communities or modules will have a high $Q$, whereas networks with little community structure (such as random networks) will have a low $Q$. We used the Brain Connectivity Toolbox's [51] `modularity_und.m` function, which implements Newman's spectral community detection algorithm [52], to compute network modularity.

To directly study the relationship between network efficiency, modularity, and integrated information, we followed the network generation procedure introduced by Watts and Strogatz in their canonical paper on small-world networks [53]. In their paper, Watts and Strogatz begin with completely regular lattice networks, in which nodes are only connected to their neighbors; they then systematically increase a parameter $p$, which is the probability that a given node will re-wire a local connection and connect to any random node in the network. A $p$ of 0 yields a completely regular lattice network, a $p$ of 1 yields a completely random network, and intermediate values of $p$ yield "small-world" networks, which are highly clustered like regular lattice networks but also have short characteristic path lengths like random networks (Fig 5A). The parameter $p$ also systematically controls the global efficiency of the network: higher values of $p$ produce networks with higher global efficiency [49] (Fig 5B). We also show that $p$ systematically decreases network modularity (Fig 5C).

Since up until this point we have only shown that our spectral clustering-based approach can find the MIB of brain-like networks of coupled oscillators, autoregressive signals generated from brain-like networks, and in real brain data, we first checked whether spectral clustering can also find the MIB in small lattice networks, small-world networks, and random networks

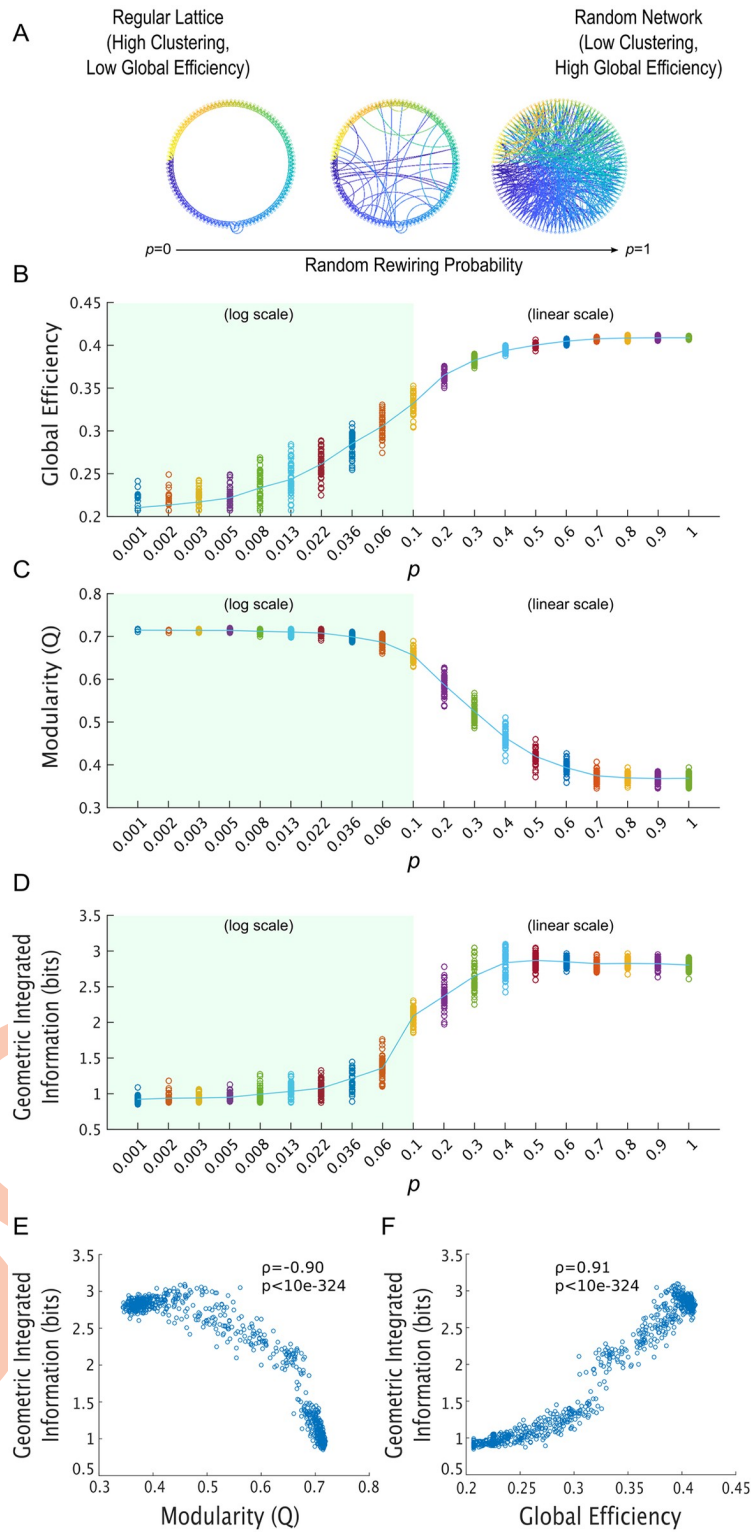

**Fig 5. The method presented in this paper for quickly identifying a network's MIB using spectral clustering makes it possible to quickly measure integrated information in large brain networks.** A straightforward first-pass at an application for our method is to evaluate the long-held and untested assumptions that the "global efficiency" of a network reflects its capacity for information integration and that the modularity of a network underpins the segregation of information. **A** Following the procedure introduced by Watts and Strogatz [53], we systematically

increased the global efficiency of our networks by increasing their rewiring probability $p$. Following Watts and Strogatz [53], we varied $p$ on a log-scale between 0.001 and 0.1; to explore the full parameter space, we also linearly varied $p$ between 0.1 and 1. For each value of $p$, we generated 50 100-node networks, and generated time-series data for each of those networks using the stochastic Rössler oscillator model. We then used our spectral clustering-based technique to measure geometric integrated information in these networks. **B** As expected [49], increasing $p$ increased the global efficiency of the networks. Here, each dot corresponds to the global efficiency of one network of coupled Rössler oscillators with that particular value of $p$. The green line passes through the mean across networks. **C** Increasing $p$ also systematically *decreased* the modularity $Q$ of the networks. **D** A higher probability $p$ of forming long-distance network connections, which increases global efficiency, led to higher integrated information (*non-normalized*). **E** There was a strong negative correlation between the networks' structural modularity and how much information they integrate, in bits (Spearman's $\rho$ = -0.90, $p < 10^{-324}$). Note that the gap around $Q = 0.65$ occurs at the transition from the log variance of $p$ to the linear variance of $p$ (**C**). **F** There was a strong positive correlation between the networks' global efficiency and how much information they integrate, in bits (Spearman's $\rho$ = 0.91, $p < 10^{-324}$). Note that the gap around $E = 0.32$ occurs at the transition from the log variance of $p$ to the linear variance of $p$ (**B**). These results support the hypothesis that network modularity supports the segregation of information, while global efficiency supports the integration of information.

of coupled oscillators. Consistent with our earlier results, we found that spectral clustering found the exact MIB (determined through a brute-force search) in almost all 14- and 16-node Rössler oscillator networks of these types that we tested (S6 Fig). As such, we felt confident that it would also give us accurate estimates of integrated information in large networks of these types. We therefore iterated through 19 values of $p$: the first 10 values were logarithmically spaced between 0.001 and 0.1 (following [53]), and the following nine values were linearly spaced between 0.1 and 1. For each value of $p$, we created 50 100-node networks, which all had the same number of edges and a mean degree of 6, and ran the Rössler oscillator model on those networks to produce 25,000 time-points of oscillatory signals. To ensure that any differences in integrated information in the resulting network dynamics were attributable to network connectivity rather than coupling strength, we set the oscillators' coupling parameter to 0.25 for all networks in this analysis (rather than determine the coupling strength through a master stability function, as we do elsewhere—see Methods).

We found that, as predicted by work in neural connectomics [37, 48, 50, 51], networks' global efficiency was tightly coupled to their capacity for information integration. Increasing the rewiring probability $p$ systematically increased both a network's global efficiency (Fig 5B) and how many bits of information are integrated across that network (Fig 5D), and decreased the networks' structural modularity. Interestingly, both global efficiency and integrated information reach a plateau around $p = 0.4$, though it is unclear from our present results why this is the case. Finally, when looking across all networks, there was a strong and significant correlation ($r = 0.91$, $p < 10^{-324}$) between the networks' global efficiency and how much information they integrate (Fig 5D) and a strong and significant anti-correlation ($r = -0.90$, $p < 10^{-324}$) between the networks' structural modularity and how much information they integrate (Fig 5E). This supports the widely held hypothesis that global efficiency determines how many bits of information a network can integrate and that modularity limits information integration, at least in the case of coupled oscillator networks. It would be interesting to see whether this relationship between network efficiency and integrated information extends to systems with non-Gaussian dynamics—a possibility we hope to explore in future work.

## Run time analysis

The results reported thus far show that our spectral clustering-based approach can accurately approximate the MIB of a system from time-series data. As a final analysis, we show that it is also much faster to run than either a brute-force search or the Queyranne algorithm for large systems, since its run time scales much less steeply (Fig 6). We simulated 25,000 time points of data using our Rössler oscillator model and artificial brain-like networks (see Methods) ranging from 10 to 120 nodes in size. We estimated integrated information using a brute-force search for the MIB in the 10- to 18-node networks, used the Queyranne algorithm for networks of 10- to 50-nodes in size, and used our spectral clustering approach for all network sizes. We empirically measured how long it took to run each of these algorithms on Matlab, using a 64-bit linux CentOS. In Fig 6 we plot the average run time across five samples of each network size. We found that, as expected, the run time for the brute-force search for the MIB scales super-exponentially; we further found that the run time for our approach scales much less steeply than does the run time for the Queyranne algorithm, which means that our method is not only more accurate than the Queyranne algorithm in finding bipartitions that minimize normalized integrated information, but is also much faster for large systems. That said, we again emphasize that the Queyranne algorithm is a valid and fast option for minimizing *non*-normalized integrated information [21].

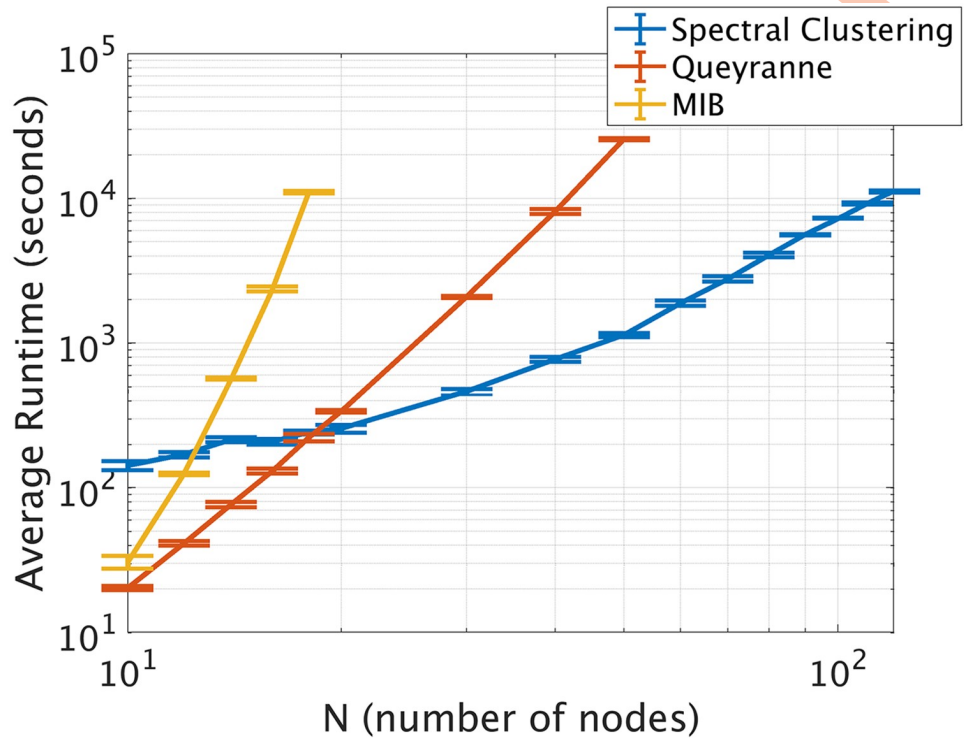

**Fig 6. Average run time across for the three algorithms as a function of network size.** Error bars indicate standard error of the mean across five networks of coupled oscillators of a given size. For very small brain-like networks (10-14 nodes), our spectral clustering-based approach is slower than either the Queyranne algorithm or a brute-force search for the MIB. This is because our algorithm searches through a fixed number of candidate graph cuts (see Methods). But, this feature is also the algorithm's strength: because our algorithm searches through the same number of candidate partitions for large systems as it does for small systems, its computation time scales much less steeply than that of the other two algorithms. If our algorithm were to search through more partitions (for e.g. by iterating through more threshold values of the correlation matrices—see Methods), then it would be slower, but its run time would still scale far less steeply than the other two algorithms, because the number of candidate partitions would remain fixed.

## Discussion

We have presented in this paper a method for measuring integrated information in large systems, using time-series observations from those systems. Specifically, we presented a robust approximate solution to the search for the minimum information bipartition of large networks, a problem that has impeded efforts to measure integrated information in large brain networks. Our proposed method for quickly partitioning brain networks to find the MIB is drawn from well-established methods in neuroimaging (for a recent review of the use of graph clustering on neural correlation matrices to identify functional sub-networks of the brain, see [22], and for the specific use of spectral clustering in such analyses, see [35, 38]). Although the Queyranne algorithm has previously been shown to successfully find bipartitions that minimize *non*-normalized integrated information [21], the algorithm usually finds one-vs-all network partitions, even when trying to find a partition that minimizes *normalized* integrated information (as we report here). That said, we agree with Kitazono and colleagues [21] that it would be fruitful to consider methods that combine our spectral clustering-based approach with their Queyranne algorithm-based approach.

It is worth pointing out that although spectral clustering found the MIB or partitions close to the MIB in the majority of both real and simulated signals for which the ground-truth MIB could be computed, it did not always yield perfect results. While it is still unclear what conditions ensure that spectral clustering will find the exact MIB, we note that in the analyses performed here, the performance of spectral clustering was correlated with the strength of interactions between units separated by the spectral partition (S4 Fig).

Importantly, our solution passed a number of basic but challenging tests involving artificial and real brain recordings. As a first application of our result, we investigated the relationship between integrated information and network structure. We found that, consistent with earlier predictions [37, 48, 50, 51, 54], networks with a high global efficiency produce high integrated information and that networks with high structural modularity produce low integrated information (Fig 5). This observation may help in pinpointing brain structures with high levels of information integration. For example, it has been assumed that the cerebellum does not integrate much information because of its highly modular architecture, while the rich, recurrent cross-module connectivity of the thalamocortical system has been assumed to allow for high levels of information integration [55–57]. Our simulation-based results support this hypothesis, though the truth of the matter will clearly need to be determined on the anvil of experiment.

We also found that our method for identifying the MIB of large systems split posterior sensory areas from anterior association areas in both monkey cortices we tested (Fig 4). In strict mathematical terms, this means that activity in posterior and anterior regions evolved largely independently over time. We note that both monkeys were awake and resting while the data we analyzed were collected; it would be interesting to see whether the demarcation of independent information-processing sub-networks might vary as a function of cognitive task or brain state.

Because our solution to the problem of searching for the MIB in large networks has made it possible to measure integrated information in real brains, we envision the described solution becoming a broadly applicable tool for neuroscience. In particular, our solution can help to elucidate the function of recurrent brain networks, just as the information-theoretic measure of channel capacity revealed coding schemes in feedforward brain circuits [2–8]. Our method can also be used to directly test the Integrated Information Theory of Consciousness [13], for example by measuring changes in information integration during states of unconsciousness, like anesthesia. With respect to the applicability of our method to the Integrated Information Theory of Consciousness, it is worth pointing out one fascinating result here, which was that in the macaque brains, integrated information peaked at a time-lag of around 100 ms (S3 Fig), which roughly corresponds to the observed timescale of conscious human perception [58, 59]. This matching of time scales is one prediction of the Integrated Information Theory of Consciousness [13, 28], though this correspondence should be investigated more systematically in future empirical work.

Given the potential usefulness of measuring integrated information in complex systems more generally, our method may also be of use to researchers in other fields as well. To facilitate such research, we have made our Matlab toolbox publicly available.

## Methods

### Simulating connectomes

We here describe our algorithm for generating artificial brain-like networks or "connectomes". First, following insights from the evolutionary neuroscience literature [60], the number of modules in our networks was equal to the log of their number of nodes, rounded up. The sizes of the modules in these networks were random, though the sizes of the modules did not vary significantly because each node had an equal probability of being assigned to any given module. Undirected edges were cast between nodes according to two different probabilities: for a pair of nodes $i$ and $j$ where $i \neq j$, an edge was cast between $j$ and $i$ according to a probability $p_{int}$ if both nodes were in the same module and with probability $p_{ext}$ if they were in different modules. For a given network with $M$ modules and for a given module with $n$ nodes, if $n \geq 4$, then $p_{int} = \frac{4.5}{n}$ and $p_{ext} = \frac{3.3}{nM}$; otherwise $p_{int} = \frac{4}{n}$ and $p_{ext} = \frac{3.75}{nM}$.

To mimic a basic Hebbian process, the nodes that made the most connections were then rewarded with even more connections and the nodes that made the fewest connections were punished by having their connections pruned. The process works like so: after edges have been cast according to the two probabilities $p_{int}$ and $p_{ext}$, find $q$, such that around 38% of nodes have made fewer than $q$ connections (this parameter of 38% was chosen somewhat arbitrarily, but it reliably led to a log-normal degree distribution as desired). Create a vector $x$ with elements $[q - 1, q, q + 1, \ldots, f + 5]$, where $f$ is the largest number of connections that any node in the network made in the previous step of casting out connections. Create a second vector $y$ of the same length as vector $x$. The first $\frac{l}{4}$ elements of $y$ are set to 1, and the last $l - f + 1$ elements of $y$ are set to $Z$, where $Z = \sqrt{N} + \log \frac{N}{7}$, $N$ is the number of nodes in the network, and $l$ is the length of vectors $x$ and $y$. The middle $w$ elements of $y$, where $w = f - \frac{l}{4} + 1$, are replaced with the vector $\left[1, 1 + \frac{Z}{w}, 1 + 2\frac{Z}{w}, \ldots, Z\right]$. A sigmoid function $S$ is fit to $x$ and $y$. For every node in the network, random connections are pruned or added, such that every node now has $S(c)$ connections, where $c$ is the number of edges the node had before pruning or adding connections. All networks were checked to ensure that in a given network, any node could be reached by any other node. The resulting networks recapitulated basic features of brain networks, including a modular structure with rich cross-module connectivity [22], as well as a log-normal degree distributions with long right tails [44].

### Simulating time-series data with coupled stochastic Rössler oscillators

To simulate oscillatory brain signals from our artificial networks, we used a stochastic Rössler oscillator model. We chose to simulate data using Rössler oscillators because, as has been previously shown [43], they follow a multivariate normal distribution when weakly coupled (S2 Fig). The system of Rössler oscillators is modeled by the following differential equations:

$$\dot{x}^i = -wy^i - z^i - \sigma \sum_{h=1}^{N} g_{ih} x^h \tag{13}$$

$$\dot{y}^i = wx^i + ay^i + d\eta^i \tag{14}$$

$$\dot{z}^i = b + (x^i - c)z^i \tag{15}$$

where, following previous literature [43, 61], $a = 0.2$, $b = 0.2$, and $c = 9$. The oscillation frequencies $w$ were normally distributed around a mean of 10 with a standard deviation of.1. $d$ was set to 750, and $\eta^i$ is Gaussian noise. $g_{ih}$ are the coefficients of the network's Laplacian matrix, and $\sigma$ is the coupling strength between oscillators. For all simulations other than the ones reported in Fig 5 (where the coupling was 0.25 for all networks), $\sigma$ was determined using a master stability function. Master stability functions give the lower and upper bounds for the coupling strengths that ensure network synchronizability. For networks of coupled Rössler oscillators, the lower-bound for the coupling strength is 0.186 divided by the second top eigenvalue of the network's Laplacian matrix, and the upper-bound is 4.614 divided by the last eigenvalue of the network's Laplacian matrix [62, 63]. For each network, $\sigma$ was set to the half-way point between these lower- and upper-bounds. The equations were integrated with a Euler algorithm, with $dt = 0.001$. For our time-series, we took the $y$ component of these equations, which yielded rich synchronization dynamics and followed a multivariate normal distribution (S2 Fig).

## Reducing the search space for the MIB

As shown in [23], spectral clustering provides an approximate but robust solution to the "normalized cut" or Ncut problem in graph theory. The problem is motivated by a body of work on how to partition a graph $G = (V,E)$, with $V$ vertices and $E$ edges, into disjoint subsets $A$, $B$, $A \cup B = V$, $A \cap B = \emptyset$. The Ncut problem entails finding a network cut which minimizes the following measure:

$$Ncut(A, B) = \frac{cut(A, B)}{assoc(A, V)} + \frac{cut(A, B)}{assoc(B, V)} \tag{16}$$

where $cut(A, B)$ is the sum of edges (binary or weighted) crossing a particular cut, $assoc(A, V)$ is the sum of edges between community $A$ and the entire network, and $assoc(B, V)$ is similarly the sum of edges between community $B$ and the entire network. Dividing $cut(A, B)$ by the normalization factors $assoc(A, V)$ and $assoc(B, V)$ helps ensure that the clusters separated by the bipartition are relatively balanced in size, and as such serves the same function as the normalization function $K$ (Eq 10) in the search for the MIB.

Shi and Malik [23] developed a fast spectral clustering algorithm that can quickly find a partition that (approximately but robustly) minimizes the Ncut function. The algorithm applies $k$-means clustering to the eigenvectors corresponding to the top $k$ eigenvalues of a network's Laplacian matrix, where $k$ is the number of communities being split (so, for a bipartition, $k = 2$). Though many other clustering methods are available, we chose spectral clustering because it is particularly well-suited for *normalized* clustering problems, and as such is appropriate for the search for the MIB.

The principle contribution of this paper is the empirical finding that the MIB of a network can be approximated by applying spectral clustering to correlation matrices of time-series data. To get a range of candidate partitions from a single correlation matrix, we first applied a power adjacency function [36] to the correlation matrix $C$, such that every correlation value $r_{ij}$ in $C$ is mapped onto a continuous edge weight $w_{ij}$:

$$w_{ij} = \left(\frac{r_{ij} + 1}{2}\right)^\beta \tag{17}$$

The value chosen for $\beta$ determines the shape of the power adjacency function. We iterated through 10 values of $\beta$, logarithmically spaced between 1 and 10. For every resulting power adjacency transformation of $C$, we then iterated through a range of cutoff values (from the 0th to the.99th percentile of weights in steps of 0.005), and for every iteration, all edge weights less than that cutoff value were set to 0 (following [37–41]). Spectral clustering was then applied to the Laplacian matrix computed from each adjacency matrix, as well as to the Laplacian matrix computed from the un-thresholded correlation matrix. In total, this resulted in 2189 candidate partitions for each dataset. $\Phi^G$ (normalized) was calculated for each of these candidate partitions, and we chose among these the partition that minimized $\Phi^G$ (normalized) as our spectral clustering-based alternative to the MIB (identified through a brute-force search). Note that, to our knowledge, there is no analytic guarantee that the MIB will be among these 2189 candidate partitions, and so the work presented here can be seen as a numerical experiment strongly motivating the proposal that there is a relationship between the MIB and the spectral partition of the correlation matrix of time-series data. In future work, we hope to analytically study this relationship in greater depth.

## ECoG preprocessing

ECoG data from the left cortex of two monkeys, Chibi and George, is publicly available on neurotycho.org [46]. Data from 128 electrodes were available for over an hour of recording from both monkeys. We selected the first 50,000 ms of data from both monkeys. The data were then down-sampled to 500 Hz, demeaned, de-trended, and band-stop filtered for 50 Hz and harmonics, which is the line noise in Japan (where the data were collected). Data were then re-referenced to the common average across electrodes. We then visually inspected the data for artifacts. Segments of data with artifacts that spread across more than one electrode were removed from all electrodes, and individual electrodes with consistent artifacts that did not spread to their neighbors were removed entirely (electrodes 14, 28, and 80 were removed for George, and electrodes 17, 53, and 107 were removed for Chibi). The pre-processed ECoG data were approximately multivariate normal (S2 Fig), allowing for the fast measurement of integrated information.

## Supporting information

**S1 Text. Extrapolating integrated information to infinite observations.**
(PDF)

**S1 Fig. Results of extrapolation analysis.**
(PDF)

**S2 Fig. Assessing data multivariate normality.**
(PDF)

**S3 Fig. Integrated information as a function of time-lag.**
(PDF)

**S4 Fig. Integrated information and partition similarity to the MIB.**
(PDF)

**S5 Fig. Cluster correlation as a predictor of MIB estimation accuracy.**
(PDF)

**S6 Fig. Spectral clustering accuracy in Watts-Strogatz networks.**
(PDF)

**S7 Fig. Spectral clustering accuracy in small autogregressive systems.**
(PDF)

**S8 Fig. Spectral clustering accuracy in large autogregressive systems.**
(PDF)

**S9 Fig. Surrogate analysis of monkey ECoG results.**
(PDF)

## Acknowledgments

We thank Nihat Ay and Daniel Lurie for their feedback on earlier versions of this manuscript.

## Author Contributions

**Conceptualization:** Daniel Toker, Friedrich T. Sommer.

**Data curation:** Daniel Toker.

**Formal analysis:** Daniel Toker.

**Funding acquisition:** Daniel Toker.

**Investigation:** Daniel Toker.

**Methodology:** Daniel Toker, Friedrich T. Sommer.

**Project administration:** Friedrich T. Sommer.

**Software:** Daniel Toker.

**Supervision:** Friedrich T. Sommer.

**Validation:** Daniel Toker.

**Visualization:** Daniel Toker.

**Writing – original draft:** Daniel Toker.

**Writing – review & editing:** Daniel Toker, Friedrich T. Sommer.